%% file: eros.tex
\documentclass[12pt,twoside,a4paper]{article}
\usepackage{epsfig}
\usepackage{setspace}

\input macropere2006.tex

\begin{document}


\Title{ACCESSING THE INNERMOST REGIONS OF ACTIVE GALACTIC NUCLEI}

\begin{raggedright}


E.~Ros
\footnote{ER has been partly supported by Grant AYA2005-08561-C03-03 of the Spanish DGICYT to attend this conference.}\\

{\small{\it Max-Planck-Institut f\"ur Radioastronomie,
Bonn, GERMANY}} \\

\medskip

M.~Kadler
\footnote{MK has been supported by the International Max Planck Research School
for Radio and Infrared Astronomy at the Universities of Bonn and Cologne
and by a NASA Postdoctoral Program Fellowship appointment conducted
at the Goddard Space Flight Center.}\\

{\small{\it 
Max-Planck-Institut f\"ur Radioastronomie,
Bonn, GERMANY
and\\
NASA/Goddard Flight Space Center, 
Greenbelt, MD
, USA
}} \\



\runninghead {E.~Ros \& M.~Kadler}
{Accessing the innermost regions of AGN}

\thispagestyle{empty}
\medskip\smallskip


{\small 
{\bf Abstract:} 
Very-long-baseline interferometry
can
image the parsec-scale structure of radio jets, but the accretion
disk close to the black hole remains invisible.  One way to probe
this accretion flow is provided by 
X-ray flux density monitoring and spectroscopy.
Here we report on preliminary results of a multi-band campaign
on NGC\,1052 with the goal of combining both approaches to access
to the innermost regions of this active galaxy and to establish
a connection between the relativistic jets and the accretion region.
}

\end{raggedright}
\begin{singlespace}


\vspace{-5pt}
\paragraph{Introduction}
Active galactic nuclei (AGN) can be studied throughout the electromagnetic
spectrum.  Over decades, the structure of their jets at parsec
scales has been studied via very-long-baseline interferometry (VLBI).
Detailed X-ray monitoring has become
possible in the last years with satellites like
\textit{RXTE}, \textit{XMM-Newton}, and
\textit{Chandra} and has yielded not only images at kiloparsec scales, but
also spectra and brightness variability on short time-scales.  The 
relativistic Fe K$\alpha$ line at $\sim$6.4\,keV probes directly the 
accretion flows (usually found in Seyfert~1 galaxies;
Nandra et al. \cite{nandra97}).  This feature had not been observed
in AGN until recently.

An extensive study of AGN is being performed at radio frequencies since
the mid 1990s, namely the VLBA\footnote{The Very Long Baseline Array
is operated by the National Radio 
Astronomy Observatory, a facility of the USA National Science Foundation
operated under cooperative agreement by Associated Universities, Inc.}
2\,cm Survey (see Zensus et al. \cite{zensus02} and Kellermann et al.\ 
\cite{kellermann04})),
continued as MOJAVE since 2002 (see Lister \& Homan
\cite{lister05}).  This observational effort consists of the monitoring 
of up to 200 sources to determine the kinematics of their milliarcsec-scale
jet systems.  
A sub-sample has
been studied at X-ray wavelengths by Kadler \cite{kadler05} with 
the aim of comparing
different physical properties of the central regions and of the
extended jets and eventually to connect the phenomena observed
in the radio jets with the accretion disk changes seen at X-ray wavelengths.  
The majority of radio-loud, core-dominated AGN is well described by a one- or two-component power law, with photon indices typically around 1.6 to 1.7.
A small subset exhibits additional X-ray spectral features that allow probing
the inner accretion flow.


\vspace{-5pt}
\paragraph{Multi-band observations of NGC\,1052}
%
The radio loud, core-dominated active galaxy NGC\,1052 ($z$=0.0049)
has provided key results in the context mentioned above.
This source has been monitored for more than a decade as part of the 
VLBA 2\,cm survey/MOJAVE projects.  Multiple sub-parsec scale 
features display outward motions of 0.26$c$
in the jet and the counter-jet (e.g.\ Vermeulen et al. \cite{vermeulen03}).
At X-ray wavelengths, the source presents an unusually flat spectrum 
with a soft excess
(e.g.\
Weaver et al. \cite{weaver99}, Kadler et al. \cite{kadler04}) 
Kadler \cite{kadler05} reported a relativistic broad iron line in
NGC\,1052, the first such feature found in a radio-loud AGN.  The line
profile discloses the disk nature and 
shows variability from epoch 2000.03 to 2001.62.
A major jet-production event took place around the same time.

We have set up a radio/X-ray monitoring of NGC\,1052\footnote{%
The NGC\,1052 campaign is carried out in collaboration with
J.A.\ Zensus, E.\ Angelakis, K.A.\ Weaver, J.\ Kerp, S.\ Kaufmann, 
A.P.\ Marscher, H.D.\ Aller, M.H.\ Aller, J.I.\ Irwin, and Y.Y.\ Kovalev.}
to trace the ejecton of new components right
from the jet base along both sides of the jet system with the VLBA,
together with X-ray monitoring with \textit{RXTE} 
to probe the variability and spectrum
of the accretion disk and eventually establish correlations
and cause-effect relationships.  The campaign is described in 
Table~\ref{table:table-ros}.  First results are described here.

\begin{figure}[p]
\begin{center}
\epsfig{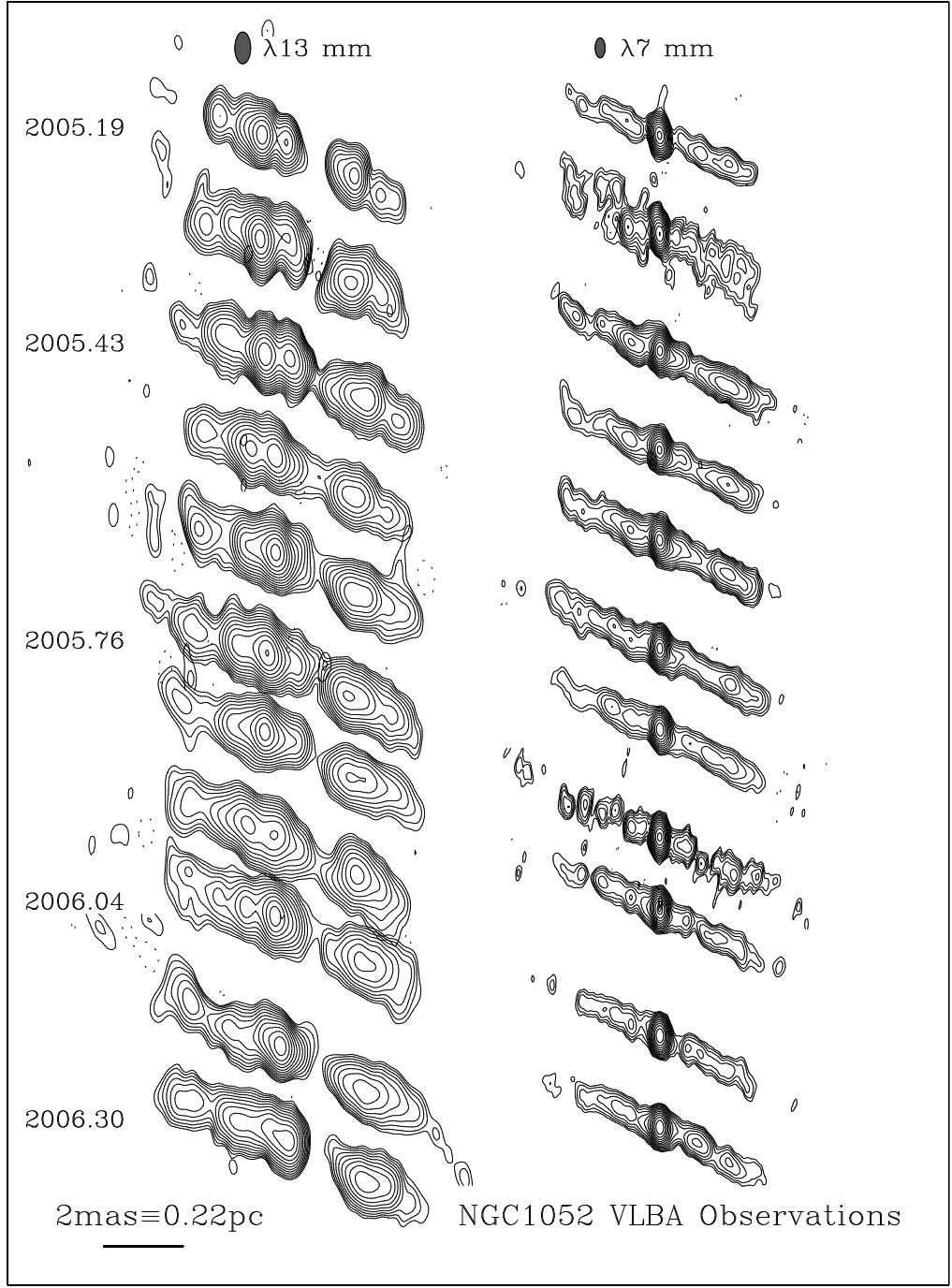}
\caption{The varying radio structure of NGC 1052 at sub-parsec scales during our monitoring program.  The images are shown spaced by their relative time intervals.  
The common restoring beams are 0.8$\times$0.4\,mas$^2$ 
and 0.5$\times$0.25\,mas$^2$,
respectively (both at
P.A.\ 0$^\circ$).
The images have been aligned arbitrarily to the gap between jet and counter-jet (left column) and at the brightness peak (right column).  A detailed phase-referencing analysis is pending.  }
\label{fig:ros}
\end{center}
\end{figure}
%

\begin{table}[t!]
\begin{footnotesize}
\begin{center}
\begin{tabular}{@{}p{1.9cm}p{3.3cm}p{7.8cm}@{}}  
Facility & Obs. type & Comments \\ \hline
\textit{RXTE} & 2--10\,keV monitoring & 30 epochs of 2\,ks each, scheduled every three weeks\\
\textit{Chandra} & Imaging/Spectroscopy & One deep obs.\ in September 2005 \\
\textit{XMM-Newton} & Imaging/Spectroscopy & One triggered obs.\ in Feb 2006 \\ \hline
Effelsberg & $\lambda\lambda$13/6/3.6/2.8/2/1.3/ 0.9\,cm\,light\,curve & $\sim$70\,h obs.\ scheduled every three weeks \\
UMRAO \& RATAN-600 & $\lambda\lambda$31/13/7.7/6/3.9/3.6/ 2.7/2/1.4\,cm\,light\,curve & Observations inserted in long-term programs \\
VLBA & $\lambda\lambda$7/13\,mm imaging & 18 epochs of 6\,h each scheduled every six weeks \\ \hline
\end{tabular}
\caption{Multi-mission campaign on NGC\,1052}
\label{table:table-ros}
\end{center}
\end{footnotesize}
\end{table}

The results from the 
\textit{RXTE} light curve show that the historically well 
known ``unusually flat X-ray spectrum'' of
NGC\,1052 is seen only in individual epochs. At other times, the spectrum is steeper and Seyfert like. 

The source was in a relatively bright radio state through 2005 and 
exhibited several local maxima, corresponding to 
ejections of new components into the jet and counter-jet system. 
VLBI imaging results 
at 7\,mm and 13\,mm wavelengths are presented in Fig.~\ref{fig:ros}.
The images at 7\,mm show the emission in the region between the jet and counter jet which, at 13\,mm is a gap free of emission due to 
free-free absorption.
A detailed model fit
analysis to parametrize the kinematics and the changing flux densities
in the image features is pending, as well as the proper astrometric
registration of the images from a phase-referencing analysis.

\vspace{-5pt}
\paragraph{Summary and Prospects}
The combination of radio and X-ray observations of NGC\,1052 adresses the question of the radio loudness of AGN and its connection with jet physics.  In this context, NGC\,1052 is a key source for jet/disk connection studies, since it is favorably observed at both frequency ranges.
In general this connection can be established by studying the 
X-ray variability on weekly time-scales, relativistic iron line changes, and
the ejection of new VLBI components.

Looking to the future, the telescopes for the next decade will make possible similar detailed studies in many sources.  The Square Kilometre Array era
will open the avenue to radio studies of active galaxies in the radio-quiet regime (e.g., Seyfert galaxies with broadened iron lines), and 
\textit{Constellation-X} (see White et al.\ \cite{white04}) will complement the SKA in the X-ray band.  
It will
provide enough sensitivity to probe the accretion disk changes in almost real 
time in prominent broad-iron-line systems and will disclose many other 
radio-loud AGN with broad iron lines that are dulted by the underlying 
power-law continuum.



\begin{footnotesize}

\end{footnotesize}
\end{singlespace}
 
\end{document}

%% file: macropere2006.tex
\def\beq{\begin{equation}}
\def\eeq#1{\label{#1}\end{equation}}
\def\eeqn{\end{equation}}

\def\beqa{\begin{eqnarray}}
\def\eeqa#1{\label{#1}\end{eqnarray}}
\def\eeqan{\end{eqnarray}}

\let\bar=\overbar

\newcommand{\captionfonts}{\footnotesize}

\makeatletter  
\long\def\@makecaption#1#2{%
  \vskip\abovecaptionskip
  \sbox\@tempboxa{{\captionfonts #1: #2}}%
  \ifdim \wd\@tempboxa >\hsize
    {\captionfonts #1: #2\par}
  \else
    \hbox to\hsize{\hfil\box\@tempboxa\hfil}%
  \fi
  \vskip\belowcaptionskip}
\makeatother   

\def\Title#1{

\setbox0\vbox{\vskip-70pt
{\scriptsize{
{\em
\begin{raggedright}
Primer Encuentro de la Radioastronom{\'{\i}}a Espa{\~n}ola\\
~~~~~~~~~~~~~~~``Memorial Lucas Lara''\\ 
J.C.~Guirado, I.~Mart{\'{\i}}-Vidal, and J.M.~Marcaide (eds.)\\ 
May 9th-11th 2006, Valencia, Spain\\
\end{raggedright}
}}}
}

\wd0=0pt
\ht0=0pt
\box0

\begin{doublespace}
\vspace*{-25pt}
\begin{center} {\Large {\bf #1} } \end{center}
\end{doublespace}

\medskip

}

\newcommand{\runninghead}[2]
{\markboth{\small{\it {#1}}}{\small{\it{#2}}}}

\def\Dslash{\not{\hbox{\kern-4pt $D$}}}
\def\dslash{\not{\hbox{\kern-2pt $\del$}}}

\def\msb{{\bar{\ssstyle M \kern -1pt S}}}

\def\Dm2{\Delta m^{2}}
\def\eV2{\mbox{eV}^{2}}
\def\B8{^{8}\mbox{B}}
\def\Be7{^{7}\mbox{Be}}
